\documentclass[10pt,twocolumn,aps,showpacs]{revtex4}
\usepackage{amsfonts}
\usepackage{amsmath}
\usepackage{graphicx}

\newtheorem{theorem}{Theorem}

\newtheorem{proposition}[theorem]{Proposition}

\newcommand{\qed}{\rule{7pt}{7pt}}
\newenvironment{proof}{\noindent{\bf Proof}\hspace*{1em}}{\qed}

\def\be{\begin{equation}}
\def\ee{\end{equation}}
\def\bea{\begin{eqnarray}}
\def\eea{\end{eqnarray}}
\def\ben{\begin{eqnarray*}}
\def\een{\end{eqnarray*}}

\def\eea{\end{eqnarray}}
\def\ben{\begin{eqnarray*}}
\def\een{\end{eqnarray*}}

\newcommand{\R}{\mathbb{R}}

\newcommand{\bra}[1]{\langle #1 |}
\newcommand{\ket}[1]{| #1 \rangle}

\newcommand{\id}{\openone}
\newcommand{\hlf}{\frac{1}{2}}

\newcommand{\real}{\operatorname{Re}}

\newcommand{\stabstates}{\mathcal{C}}

\begin{document} 

\title{Using unitary operations to preserve quantum states 
in the presence of relaxation}

 \author{Benjamin Recht}
 \email{brecht@media.mit.edu}
 \author{Yael Maguire}
 \author{Seth Lloyd} 
 \altaffiliation{Department of Mechanical Engineering, 
	Massachusetts Institute of Technology, Cambridge, MA}
 \author{I.~L.~Chuang}
 \author{N.~A.~Gershenfeld}
 \affiliation{Center for Bits and Atoms, 
	Massachusetts Institute of Technology, Cambridge, MA}

\date{\today}

\begin{abstract} 
When a quantum system interacts with an external environment, it
undergoes the loss of quantum correlation (decoherence) and the loss
of energy (relaxation) and eventually all of the quantum information
becomes classical.  Here we show a general principle to use unitary
operations to establish and preserve particular non-equilibrium states
in arbitrary relaxing quantum systems.  We elucidate these concepts
with examples of state preservation in one-spin and two-spin entangled
systems.
\end{abstract}


\maketitle

Quantum systems undergo damping and decoherence when they interact
with thermodynamic systems.  Preventing such non-unitary behavior is
one of the biggest challenges facing the engineering of quantum
information technology.

Much work has been done to develop quantum error correction schemes to
correct these non-unitary errors~\cite{Shor95} \cite{Steane96}.  When
the errors are below known thresholds~\cite{Knill98a}, these schemes
can preserve quantum information for an arbitrarily long time.  On the
other hand, error correction is likely to be difficult because the
required fidelity is beyond current practice and a large supply of
ancilla qubits is needed ~\cite{Ekert96}\cite{Bennett96}\cite{Knill97}.

When the dissipation modes are purely decoherence processes,
Decoherence Free Subspaces that are unaffected by the decoherence can
be relabeled and used for storing and processing quantum
information~\cite{Zanardi97}\cite{Lidar98}.  In this paper, we will
look at the situation when the dissipation is damping or {relaxation}.
Surprisingly in this case, unitary operations can be used to preserve
a large submanifold of quantum states for an arbitrarily long time.
We will outline the geometry of such situations and then detail two
examples. The first example will illustrate the dynamics of one-spin
systems.  The second example will show how to preserve a pair of
coupled spins in a particular entangled state.

First, consider a general open quantum system of dimension $N$ with
density matrix $\rho$.  A general time evolution of this system is given
by the mapping 
\begin{equation}
	\rho \mapsto \mathcal{E}_t(\rho) = \sum_k E(t)_k \rho
		E(t)_k^\dagger 
\end{equation} 
where the $E_k$ satisfy $\sum_k E_k(t)^\dagger E_k(t)=\openone$ for
all $t$ and are called {Kraus operators}~\cite{Kraus71}.  If our open
system evolution is Markovian, and satisfies the algebraic property
$\mathcal{E}_t\mathcal{E}_s=\mathcal{E}_{t+s}$, then the evolution is
called a {quantum dynamical semigroup}.  The dynamics are completely
determined by the generator of this semigroup.  The most general
differential equation for such a generator is given by the {Lindblad
Equation}
\begin{equation}~\label{eqn:lindblad}
    \dot{\rho}=-i[H,\rho]+\frac{1}{2} \sum_{k=0}^{K}
            [L_k, \rho L_k^\dagger]+[L_k\rho,
            L_k^\dagger] 
\end{equation} 
where $H$ is a Hermitian matrix and $L_k$ are a set of $N\times N$
matrices~\cite{Lindblad76}.  If the $L_k$ are identically zero, then
the Lindblad equation reduces to a Schr\"{o}dinger equation and we
call the matrix $H$ the {Hamiltonian part} of the dynamics. In turn,
the set terms of the Lindblad equation involving the operators $L_k$
is called the {dissipative part} of the Lindblad equation.

Denote the trace zero Hermitian matrices $\mathfrak{su}(N)$ and choose
a basis, $F_k$, satisfying the orthogonality conditions $\mbox{Tr}(F_j
F_k)=\delta_{jk}$.  Every density matrix $\rho$ can be written as a
sum
\begin{equation}
    \rho=\frac{\id+\sum_{k=1}^{N^2-1} r_k F_k}{N} 
\end{equation} 
If we change our representation and define the {coherence vector},
$\vec{r}=(r_k)$, then equation~\ref{eqn:lindblad} can be written as a
standard form ordinary differential equation
\begin{equation}\label{eqn:covec}
    \dot{\vec{r}}=A\vec{r}+B\vec{r}+\vec{c}
\end{equation}
Here $A$ corresponds to the Hamiltonian part of the Lindblad
equation while $B$ and $\vec{c}$ correspond to the dissipative 
part~\cite{Alicki87}.

We will focus on a special type of quantum dynamical semigroups called
{relaxing semigroups}.  A semigroup is relaxing if for any initial
state $\vec{r}(0)$,
\begin{equation}
	\lim_{t\rightarrow \infty} \vec{r}(t)=\vec{r}_f.
\end{equation}
Such a situation occurs whenever the matrix $A+B$ is invertible and
the vector $\vec{c}$ is also nonzero.  By setting
${d\vec{r}}/{dt}=0$, we see that $\vec{r}_f=-(A+B)^{-1}\vec{c}$ is
the unique fixed point of the evolution.  If the Lindblad equation has
a unique fixed point, $\vec{r}_f$, then this fixed point is a global
attractor.  Indeed, in matrix form we find that the evolution is given
in coherence vector form by
\begin{equation}
	\vec{r}(t)=e^{(A+B)t}(\vec{r}_0-\vec{r}_f)+\vec{r}_f
\end{equation}
where $\vec{r}_f$ is the global fixed point.

Relaxing semigroups occur frequently in physical systems where the
equilibrium state of a system is known \emph{a priori}.  For example,
in a liquid state nuclear magnetic resonance experiment, the system
will always return to an equilibrium Boltzmann distribution which is
solely a function of the applied magnetic field and the temperature.
In what follows, we will show that when a quantum system evolves as a
relaxing semigroup, unitary controllers can act to stabilize a variety
of known states. 

Let us restrict attention to the following control scenario.  Suppose
a quantum system evolves as a relaxing semigroup, but that we can
apply an arbitrary controlling Hamiltonian, $H_c$, to the system, but
that we cannot adjust the dissipative terms.  In the coherence vector
representation we have
\begin{equation}
    \dot{\vec{r}}=A_c\vec{r}+B\vec{r}+\vec{c}
\end{equation}
The Hamiltonian control cannot prevent relaxation, as the eigenvalues
of the matrix $A_c+B$ will still have negative real parts, but the
following proposition shows that the controller shifts the fixed point
of the relaxing semigroup.

\begin{proposition}
Let $B$ be the dissipative part of the Lindblad equation for a
relaxing semigroup.  If $B$ is diagonalizable then for any Hamiltonian
part $A$, the matrix $A+B$ is invertible.
\end{proposition}
\begin{proof}
Let $\langle\, , \, \rangle$ be an inner product on $\R^N$.  Then the
skew symmetry of $A$ implies that $\langle \vec{r}, A\vec{r}\rangle=0$
for all $\vec{r}$.  
There exists a basis $f_j$ for $B$ such that
\begin{equation}
\begin{split}
\langle\vec{r},B\vec{r}\rangle
    &=\sum_{k=1}^N (\lambda_k +i\omega_k)\langle r,f_k \rangle^2\\
\end{split}
\end{equation}
where the $\lambda_k<0$.  The real part of the inner product is
negative for any nonzero $\vec{r}$.  This in turn means that
\begin{equation}
	\real (\langle \vec{r}, (A+B)\vec{r}\rangle) < 0
\end{equation}
for all $\vec{r}\neq 0$ which completes the proof.
\end{proof}

If we apply a Hamiltonian $A_c$ then the state
%\begin{equation}
$\vec{r}_f=(A_c+B)^{-1}\vec{c}$ 
%\end{equation} 
becomes the global attracting fixed point of our quantum system and
$A_c$ is a stabilizing controller on our system.  In particular, this
means that when $A_c$ is applied, the system's steady state is
$\vec{r}_f$, and and the system will flow to $\vec{r}_f$ independent
of the initial state.  Hence the set
\begin{equation}
	\stabstates=\{\vec{r}=-(A+B)^{-1}\vec{c} 
		\, | \, A \mbox{ is a Hamiltonian}\} 
\end{equation} 
can be made into fixed points of a relaxing semigroup using control
Hamiltonians.  Since these states will be stabilized by the dynamics,
we will refer to $\stabstates$ as the set of stabilizable states of
our semigroup.

We must note that the stabilizable states will in general be mixed
states as the length of the vector $\vec{r}_f$ will vary with the
applied Hamiltonian.  However, we will see that they can be useful for
monitoring quantum systems and for preserving entanglement.  The
following theorem describes the geometry of the set $\stabstates$.  In
the proof, we will switch between the density matrix and coherence
vector representations.

\begin{theorem}
If the fixed point, $\rho_{eq}$, of a relaxing semigroup has non-degenerate
eigenvalues, then the set of stabilizable states is a simply
connected $N^2-N$ manifold containing the fixed point of the process
and having the maximally mixed state in its closure.
\end{theorem}

\begin{proof}
Let $\rho_{eq}$ be the fixed point of the quantum process with
corresponding coherence vector $\vec{r}_{eq}$.  Consider a small
perturbation $\vec{r}=\vec{r}_{eq}+\delta \vec{r}$. It is
immediate to show that $A \vec{r}+\vec{b}=A\delta\vec{r}$.

For an infinitesimal time $\Delta t$, we have that $\rho(\Delta
t)=\rho+A\delta\vec{r}\Delta t$. If over this time, the eigenvalues of
$\rho(\Delta t)$ are the same as those of $\rho(0)$, then there exists a
unitary operator $U$ with $U\rho(t)U^\dagger=\rho(0)$ and hence $\rho$ is
the fixed point of the process
\begin{equation}
    \ldots \exp(-i H_c \Delta t) \mathcal{E}_t
    \exp(-i H_c \Delta t) \mathcal{E}_t \ldots
\end{equation}
which is generated by the Lindblad equation with $H=H_c$.

Let $\ket{\psi_n}$ be an orthonormal eigenbasis for $\rho_{eq}$
with corresponding eigenvalues $p_1 > \ldots > p_N$. We want to show
that there is an $N^2-N$ dimensional neighborhood of $\rho_{eq}$
where the eigenvalues are unchanged under such a small
perturbation. Since $\rho$ and $\mathcal{E}_{\Delta t}(\rho)$ are
perturbations of $\rho_{eq}$, we can calculate the change in the
eigenvalues
\begin{equation}
\begin{split}
\Delta p_n &= \bra{\psi_n} \rho_{eq}+\delta \rho -
                \mathcal{E}_{\Delta t}(\rho_{eq}+\delta \rho)\ket{\psi_n}\\
           &= \bra{\psi_n} \frac{A\delta \vec{r}}{N} \ket{\psi_n}
\end{split}
\,.
\end{equation}
The set of matrices $ M\in \mathfrak{su}(N)$ such that $\bra{\psi_n} M
\ket{\psi_n}=0$ has dimension $N^2-N$ as it corresponds to those
traceless Hermitian matrices with zeros on the diagonal.

If $\rho$ is in the set of stabilizable states and has a corresponding
Hamiltonian $H$, then there is a corresponding $\rho_\mu$ for $\mu H$.  
At the limit where $\mu=\infty$, $\rho_\infty=\id/N$. Hence, $\id/N$ is a
limit point of $\stabstates$.
\end{proof}

The preceding argument sets an upper bound on the dimensionality of the
space of stabilizable states.  If the fixed point has degenerate
eigenvalues or the set of controller Hamiltonians is restricted to a
subspace of $\mathfrak{su}(N)$ then $\stabstates$ will have smaller
dimension.

The utility of this formalism can be explored in a one-spin example.
Consider the process of damping to the $Z$ eigenket $\ket{\uparrow}$.
In terms of the Bloch vector, the system will relax with time constant
$\gamma_1=1/T_1$ along the $Z$-axis and decohere with time constant
$\gamma_2=1/T_2 \geq \gamma_1/2$ in the $x$-$y$ plane.

The Lindblad equation which generates such a semigroup is given by
equation~\ref{eqn:lindblad} with $H=0$, $K=1$, and
\begin{equation}\label{eqn:onespinrelax}
	L_0=\sqrt{\gamma_1}I^+
	\quad \mbox{and} \quad
	L_1=\sqrt{\frac{\gamma_2}{2}-\frac{\gamma_1}{4}} Z
\end{equation}
where $I^+=(X+iY)/\sqrt{2}$ is the raising operator.  In the form of
equation~\ref{eqn:covec} this amounts to the Bloch equations with
$A=0$,
\begin{equation}
B=\left( \begin{array}{ccc} -\gamma_2 & 0 & 0\\
                            0     & -\gamma_2 & 0\\
                            0     & 0     & -\gamma_1
                            \end{array}\right), \,
\mbox{and} \quad C=\left(\begin{array}{c} 0
\\ 0 \\ \gamma_1 \end{array} \right)
\end{equation}

Parametrizing the space of Hamiltonians in the Pauli basis gives the
controller Hamiltonian
\begin{equation}
A=\left( \begin{array}{ccc} 0 & -u_z & u_y\\
                              u_z & 0 & -u_x\\
                              -u_y & u_x & 0
                            \end{array}\right),
\end{equation}
in terms of three parameter controls $\{u_x,u_y,u_z\}$ corresponding
to rotations about the $x$,$y$, and $z$ axis respectively.  

The fixed points of the Lindblad equation are given by the equation
$(A+B)\vec{r}+\vec{c}=0$ which can can be solved to find the manifold
$\stabstates$
\begin{equation}
\frac{1}{4} = (z-\frac{1}{2})^2 + \frac{\gamma_2}{\gamma_1}(x^2 + y^2)
\end{equation}
$\stabstates$ is an ellipsoid containing both the fixed point and the
maximally mixed point as we proved earlier.  Its minor axis is
governed only by the ratio of $\gamma_1$ to $\gamma_2$.

We also find the appropriate open loop controllers to reach the state
$(x,y,z)$ on $\stabstates$ are
\begin{equation}
u_x = \tau_2 \frac{y}{z} \quad \mbox{and} \quad u_y = \tau_1 \frac{x}{z}
\end{equation}

\begin{figure}[ht] \centering
\includegraphics[width=7cm]{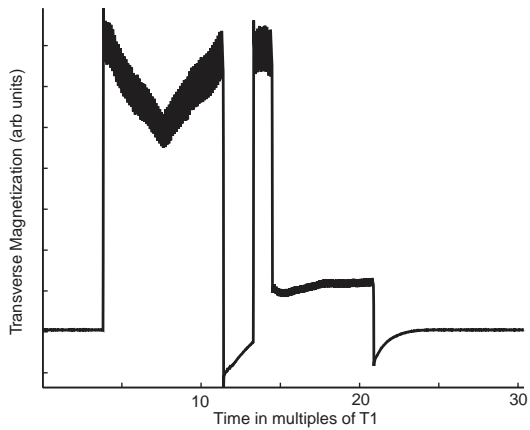} 
\caption{\label{fig:mlplot} Experimental data tracing out
the letters ``ML'' in the transverse magnetization of an NMR spin system.  
We prepared a Copper Sulfate sample in water following the prescription
in~\cite{Viola00} to create time constants $T_1=T_2=39\pm 5\mbox{ms}$.  The
signal was measured using a Varian 500 MHz NMR spectrometer. } 
\end{figure}

To ground this example in practice, let us describe how it is readily
applied applied to pulsed NMR.  We can asymptotically reach the
desired steady state by applying the unitary operations $\exp(-i (u_x
X+ u_y Y) \Delta t)$ at a repetition rate of $\Delta t$.  When $T_2$
(the transverse relaxation time $1/\gamma_2$) is comparable with $T_1$
(the longitudinal relaxation time $1/\gamma_1$), the steady state
component of the Bloch vector in the $x$-$y$ plane can be
asymptotically close to $\hlf$.  In the language of NMR, the steady
state magnetization is equal to half of the peak magnetization from a
$\pi/2$ pulse when $T_1=T_2$.  On the other hand, when $T_2 \ll T_1$,
the steady-state magnetization approaches zero. Ernst and Anderson
~\cite{Ernst66} and Freeman ~\cite{Freeman71} derived these steady
state from the Bloch equations, and our current formalism includes
their results as a special case.  Furthermore, by varying the pulse
width, and in turn the steady state, we have experimentally
demonstrated control over the NMR magnetization vector over times much
larger than T1 as shown in figure~\ref{fig:mlplot}.

Investigations into steady-state NMR on multiple spin systems has been
less broadly investigated.  It has proved successful for the specific
case of studying spin-lattice relaxation in dipolar
solids~\cite{Waugh76}\cite{Burum76}, but a general theory for multiple
spins has not been established.  Our results readily extend to higher
dimensional quantum systems, but parametrizing the set of stabilizable
states and their corresponding controllers becomes much more difficult as
the number of variables in the coherence vector scales quadratically with
the number of levels.  Nonetheless, we will demonstrate techniques for
dealing with such larger systems and describe a particular example of
using local controllers and an entangling operation to preserve a highly
entangled state.

The entanglement of a pure state, $\rho=\ket{\psi}\bra{\psi}$ of two
spin half particles is defined as
\begin{equation}
	E(\rho)=\mbox{Tr}(\rho_1 \log \rho_1)=\mbox{Tr}(\rho_2 \log \rho_2)
\end{equation}
where $\rho_j$ denotes the partial trace over the Hilbert space of the
$j$th spin.  The quantity $E$ takes values between 0 and 1 and
provides an information-like measure of the entanglement between a
pair of spins.  Correspondingly, an ``ebit'' is a unit of
entanglement.  One EPR pair has one ebit of entanglement.  For a mixed
state, we can define the entanglement of formation to be the minimum
amount of entanglement required to create this mixed states from pure
states~\cite{Bennett96}.  Precisely,
\begin{equation}
	E(\rho)=\min \sum_j p_j E(\psi_j) 
\end{equation}
where the minimum is taken over all combinations of $p_j$ and
$\ket{\psi_j}$ which yield $\rho=\sum_j p_j \ket{\psi_j}\bra{\psi_j}$.
Wootters found a functional form for this quantity which involves
extracting the eigenvalues of an algebraic function of the density
matrix~\cite{Wootters98}.  Using this metric, we show how to construct
a stabilizable state with an entanglement $0.355$.

To simplify the equations used to solve for the fixed point, we will
restrict our attention to a simple model.  Consider a two-spin system
where both spins undergo damping to the spin-up state identically and
independently with constants $\gamma_1=\gamma_2/2$.  This corresponds
to dissipative operators
\begin{equation}
	L_0=\sqrt{\gamma} I^+\otimes\id 
		\quad \mbox{and} \quad 
	L_1=\sqrt{\gamma}\id\otimes I^+ 
\end{equation}
Assume the spins are coupled via the Hamiltonian $H= J Z_1 Z_2$.  The
fixed point of this evolution is the state $\ket{\psi_0}=\ket{\uparrow
\uparrow}$.

Allow for only local Hamiltonians to be applied.  Then the admissible
Hamiltonians can parametrized as
\begin{equation}
  H_c= u_{x_1} X_1 + u_{y_1} Y_1 +u_{z_1} Z_1
	+u_{x_2} X_2 + u_{y_2} Y + u_{z_2} Z_2
\end{equation}

The coherence vector is given by $r_{jk}=\mbox{Tr}(\sigma_j \otimes
\sigma_k \rho)$ where $\sigma_j=\{\id, X,Y,Z\}$ for $j=0,1,2,3$.
Putting this all together, we get the fifteen equations
\begin{equation}\label{eqn:entsys}
	A_{jk}^{lm}r_{lm} + B_{jk}^{lm}r_{lm}+ C_{lm}=0
\end{equation}
with the coefficients of A and B readily solved for by algebra.

Consider the Hamiltonian
\begin{equation}\label{eqn:magich}
        H_c=\frac{4\sqrt{J}}{5} X_1 - J
        Z_1+\frac{4\sqrt{J}}{5}X_2 - J Z_2
\end{equation}
and let $\ket{\psi_1}=\ket{\uparrow\uparrow}$ and 
$\ket{\psi_2}=(\ket{\uparrow\downarrow}+\ket{\downarrow\uparrow})/\sqrt{2}$.  
Inverting the system in equation~\ref{eqn:entsys} and taking the
limit as $J$ approaches infinity, yields the fixed state
\begin{equation}~\label{eqn:bellfixed}
        \rho_e=\hlf 
\ket{\psi_1}\bra{\psi_1}+\hlf\ket{\psi_2}\bra{\psi_2}
\end{equation}
which indeed has the entanglement of formation of $0.355$.

\begin{figure}[ht]
\centering
\includegraphics[width=7cm]{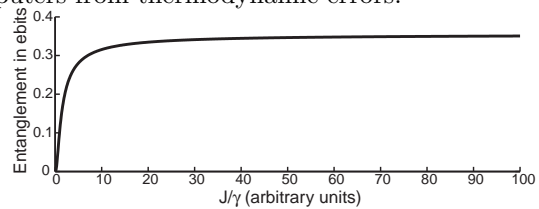} 
\caption{\label{fig:entvsJ} The entanglement of formation of the fixed
point under the Hamiltonian of equation~\ref{eqn:magich}.}
\end{figure}

The rate at which the fixed point approaches infinity is plotted in
figure~\ref{fig:entvsJ}.  Even for relatively small ratios,
$J/\gamma$, this state is close to $\rho_e$.  In the context of
quantum computation, this procedure could be used to make a ``well'' of
entanglement.  Spins that are coupled locally can be used to store a
known entangled state and then their state can be swapped into another
system which can process the entanglement for communication or
computation.

We have shown a method for analyzing relaxing semigroups and have also
shown that by applying control Hamiltonians the fixed points of these
systems can be shifted.  We have further demonstrated how to apply
these techniques to preserve known quantum states for arbitrarily long
times without the requirements of redundancy or error thresholds from
quantum error correction.

%FIGURE HERE

A framework for labeling and exploring the space of stabilizable states
in higher dimensional systems remains to be determined.  Already for
two-spins there is no intuitive description of the manifold of
stabilizable states and we have only demonstrated one example of a state
which can be stabilized.  Combining this higher dimensional labeling
with a prescription for using the stabilizable states in a coherent
fashion for quantum information processing could provide a new method for
protecting quantum computers from thermodynamic errors.

\begin{acknowledgments} 
This work was supported in part by the Center for Bits and Atoms (NSF
CCR-0122419) and the HP-MIT Alliance.  We gratefully acknowledge
Matthias Steffen, David Cory, and Aram Harrow for helpful comments
and suggestions.
\end{acknowledgments}

\bibliography{relax_101002.bib}
\bibliographystyle{apsrev.bst}

\end{document}